\documentclass[prl,twocolumn,twoside,nofootinbib]{revtex4}

\usepackage{amsmath}
\usepackage{graphicx}
\usepackage[small]{subfigure}

\newcommand{\eq}[1]{Eq.~\eqref{eq:#1}}
\newcommand{\eqs}[2]{Eqs.~\eqref{eq:#1} and \eqref{eq:#2}}
\newcommand{\fig}[1]{Fig.~\ref{fig:#1}}

\newcommand{\eg}{{\it e.g.~}}

\newcommand{\ord}[1]{{\mathcal O}(#1)}
\newcommand{\ORd}[1]{{\mathcal O}\Bigl(#1\Bigr)}

\newcommand{\nn}{\nonumber}

\newcommand{\df}{\mathrm{d}}

\newcommand{\tr}{\mathrm{tr}}

\newcommand{\al}{\alpha}

\newcommand{\ga}{\gamma}
\newcommand{\Ga}{\Gamma}
\newcommand{\de}{\delta}

\newcommand{\si}{\sigma}

\newcommand{\cG}{{\mathcal G}}
\newcommand{\cJ}{{\mathcal J}}

\newcommand{\lqcd}{\Lambda_\mathrm{QCD}}

\allowdisplaybreaks[2]

\begin{document}


\title{Fragmentation in Jets: Cone and Threshold Effects}

\author{Massimiliano Procura}
\affiliation{Albert Einstein Center for Fundamental Physics, Institute for Theoretical Physics,University of Bern, CH-3012 Bern, Switzerland\vspace{-0.5ex}}

\author{Wouter J.~Waalewijn}
\affiliation{Department of Physics, University of California at San Diego, 
La Jolla, CA 92093, U.S.A. \vspace{-0.5ex}}

\begin{abstract}
We study the fragmentation of (light) quarks and gluons to hadrons inside a jet of cone size $R$. This allows for a more exclusive analysis of fragmentation than is currently the case. The shape of semi-inclusive cross sections in the hadron energy fraction $z$ is described by fragmenting jet functions (FJFs), which we calculate in terms of $R$ and the jet energy $E$. We introduce a new joint resummation to sum the double logarithms of $R$ and $1-z$ in the FJFs, which has a similar application to initial-state radiation at hadron colliders. Our results at next-to-leading logarithmic order indicate that the resummation of the threshold logarithms of $1-z$ is already important for $z \gtrsim 0.5$ and improves the convergence of perturbation theory. Our framework may be used to analyze LHC and RHIC data and to test and tune Monte Carlo event generators.
\end{abstract}

\maketitle

\paragraph*{Introduction.}

High-energy processes with an observed hadron in the final state can be described by factorizing the short-distance (partonic) physics, which is perturbatively calculable in QCD, from (universal) non-perturbative contributions, see \eg Ref.~\cite{Collins:1989gx}. 
In this context, the (unpolarized) fragmentation functions (FFs) $D_i^h(z,\mu)$~\cite{CSfrag} encode the information on the transition from an energetic parton $i=\{g,u,\bar u, d, \dots\}$ to a
hadron $h$, which carries the fraction $z$ of its energy, plus a remainder $X$. 
The knowledge of both perturbative and non-perturbative ingredients in factorization theorems is crucial to have control on theoretical predictions. For example, a better determination of the $b$-quark FF was important for resolving a discrepancy between CDF data and theory for the $p_T$ spectrum of $J/\psi$~\cite{Cacciari:2003uh}. Furthermore, additional understanding of hadron production at high $p_T$ in $pp$ collisions is also required to determine more accurately the relative suppression of hadron spectra (jet quenching) seen in heavy-ion collisions~\cite{Agakishiev:2011dc}.

Parameterizations for the FFs have been constrained by fitting to data for single-inclusive charged hadron production in $e^+ e^-$ at next-to-leading order (NLO) in perturbation theory~\cite{fragee,Hirai:2007cx}. More recently, global analyses have been performed that also include semi-inclusive deep-inelastic scattering and/or $p p$, $p \bar{p}$ data from HERA, RHIC and the Tevatron~\cite{fragglobal}. To illustrate the current level of precision, the dominant $D_u^{\pi^+}\!(z,\mu=m_Z)$ is determined with uncertainties at the 10\% level for $z \gtrsim 0.5$~\cite{Hirai:2007cx}. The FFs of the gluon and the non-valence quarks are known even less accurately.

In contrast to the inclusive analyses listed above, the Belle collaboration is studying light-quark fragmentation in their on-resonance data, using a cut on the thrust event shape to remove the large $b$-quark background~\cite{Seidl:2008xc}. Here, the fragmentation takes place inside a (hemisphere) jet of invariant mass $s$, described by the fragmenting jet functions (FJFs) $\cG_i^h(s,z,\mu)$ that we introduced in Ref.~\cite{Procura:2009vm}. Correspondingly, $z$ should not be too small, to avoid contributions from hadrons outside the jets. For $s \gg \lqcd^2$, the FJFs can be perturbatively matched onto the FFs~\cite{Procura:2009vm}, {\it i.e.} $\cG_i^h = \sum_j \cJ_{ij}\otimes \,D_j^h$ where the convolution is in the momentum fraction $z$. In performing this OPE, we assume that parton and hadron masses \footnote{The corrections due to the hadron mass $m_h$ are of order $m_h^2/(z^2 Q^2)$, where $Q$ is the scale of the hard interaction. For light hadrons they are negligible unless $z$ is small.} are negligible compared to the jet mass. 
The Wilson coefficients $\cJ_{ij}$ describe the emissions from the parent parton at larger virtualities building up the jet while $D_j^h$ encodes physics at lower scales where hadronization effects are important.

The one-loop $\cJ_{ij}(s,z,\mu)$ were determined in Refs.~\cite{Liu:2010ng,Jain:2011xz}, and allowed us to calculate in Ref.~\cite{Jain:2011xz} the cross section for $e^+e^- \to h X$ with a cut on thrust up to next-to-next-to-leading logarithmic order in Soft-Collinear Effective Theory (SCET)\cite{SCET}. There we found correlations between the thrust cut and $z$ that are crucial for the analysis of the Belle data. Of course this analysis is subject to non-perturbative power corrections, which can be sizable at the Belle energy, and a comparison with data is desirable to assess their importance.

Here we study fragmentation inside cone jets, defined with a cone jet algorithm.  
For simplicity we discuss the case of $e^+e^- \to N$ jets, but our method can also be applied to $pp$ collisions 
\footnote{For $pp$ collisions the jet algorithms are defined in terms of $\phi$ and the pseudo-rapidity $\eta$ and thus not azimuthally symmetric around the jet axis, introducing an additional 
hurdle. We assume the cancellation of Glauber gluons in factorization theorems.}.
Following Refs.~\cite{Ellis:2009wj,Ellis:2010rwa,Jouttenus:2009ns}, the cone size is denoted by $R$ and a cutoff $\Lambda$ is applied on the energy in the region between jets \footnote{$\Lambda$ may also be defined as a jet energy veto, which is equivalent up to ${\cal O}(\alpha_s)$.}. The jets are required to be energetic and well-separated:
\begin{equation}
\tan^2 (R/2),\quad \frac{\tan^2 (R/2)}{\tan^2 (\psi/2)},\quad \frac{\Lambda}{E_\text{min}} \ll 1
\end{equation}
where $\psi$ is the minimum angular separation between jets and $E_\text{min}$ the minimum jet energy. Additional power corrections due to the jet algorithm are suppressed if $R$ is not too small~\cite{DMS}, see also~\eq{GtoD}. 
SCET can then be applied to separate hard, collinear and soft dynamics, leading to factorization formulae as in Refs.~\cite{Ellis:2009wj,Ellis:2010rwa}. Schematically, at leading power,
\begin{equation} \label{eq:fact}
  \frac{\df \si^h}{\df z}(E,R)\!= \!\int\! \df \Phi_N \tr[H_N S_N]\, \cG_i^h(E,R,z,\mu) \prod_\ell J_\ell
  \, .
\end{equation}
The hard function $H_N$ describes the hard collision and the soft function $S_N$ the soft radiation (both are matrices in color space). For each of the jets there is a jet function $J_\ell$ describing the final-state collinear radiation, which is replaced by a FJF for the jet in which the hadron is observed~\cite{Procura:2009vm}. The phase-space is denoted by $\df \Phi_N$.
The dependence on the renormalization scale $\mu$ cancels in the cross section up to the order that one is working and is used to estimate the uncertainty from higher-order corrections. 
Large logarithms of ratios of mass scales are summed by evaluating each of these functions at their natural scale, where they contain no large logarithms, and by running them to a common scale $\mu$ using their renormalization group equations. The FJFs with a cone restriction depend on $E$ and $R$ rather than $s$, and in the next section we calculate their matching onto FFs at NLO, 
\begin{align} \label{eq:GtoD}
\cG_i^h(E,R,z,\mu) &= \sum_i \int_z^1\! \frac{\df z'}{z'} \cJ_{ij}(E,R,z',\mu) D_j^h\Big(\frac{z}{z'},\mu\Big) 
\nn \\ & \quad \times
\Big[1 + \ORd{\frac{\lqcd^2}{4E^2\tan^2 (R/2)}}\Big]
\,.\end{align}
To avoid large non-perturbative corrections, $R$ should thus not be too small
\footnote{From Ref.~\cite{DMS} non-perturbative effects are expected to scale like $1/R$ for small $R$, which is consistent with the leading non-perturbative correction to the soft function.}. 
 Since the other ingredients of the factorization theorem in \eq{fact} do not affect the fragmentation variable $z$, the shape of the cross section in $z$ is completely determined by the cone FJF. (This is not the case when invariant masses are measured, as in Ref.~\cite{Jain:2011xz}, because invariant masses also receive a contribution from soft radiation.) 
The soft function is sensitive to two different scales~\cite{Ellis:2010rwa} leading to non-global logarithms~\cite{ngl} but this only affects the normalization of the cross section. 
Furthermore, we would like to mention that the generalization of our framework to transverse-momentum-dependent and polarized FFs, can be applied as a tool to obtain information on the proton spin structure 
by studying azimuthal asymmetries for hadron distributions inside a high-$p_T$ jet 
in transversely polarized $pp$ collisions~\cite{spin}.

\begin{figure*}
\includegraphics[width=0.328\textwidth]{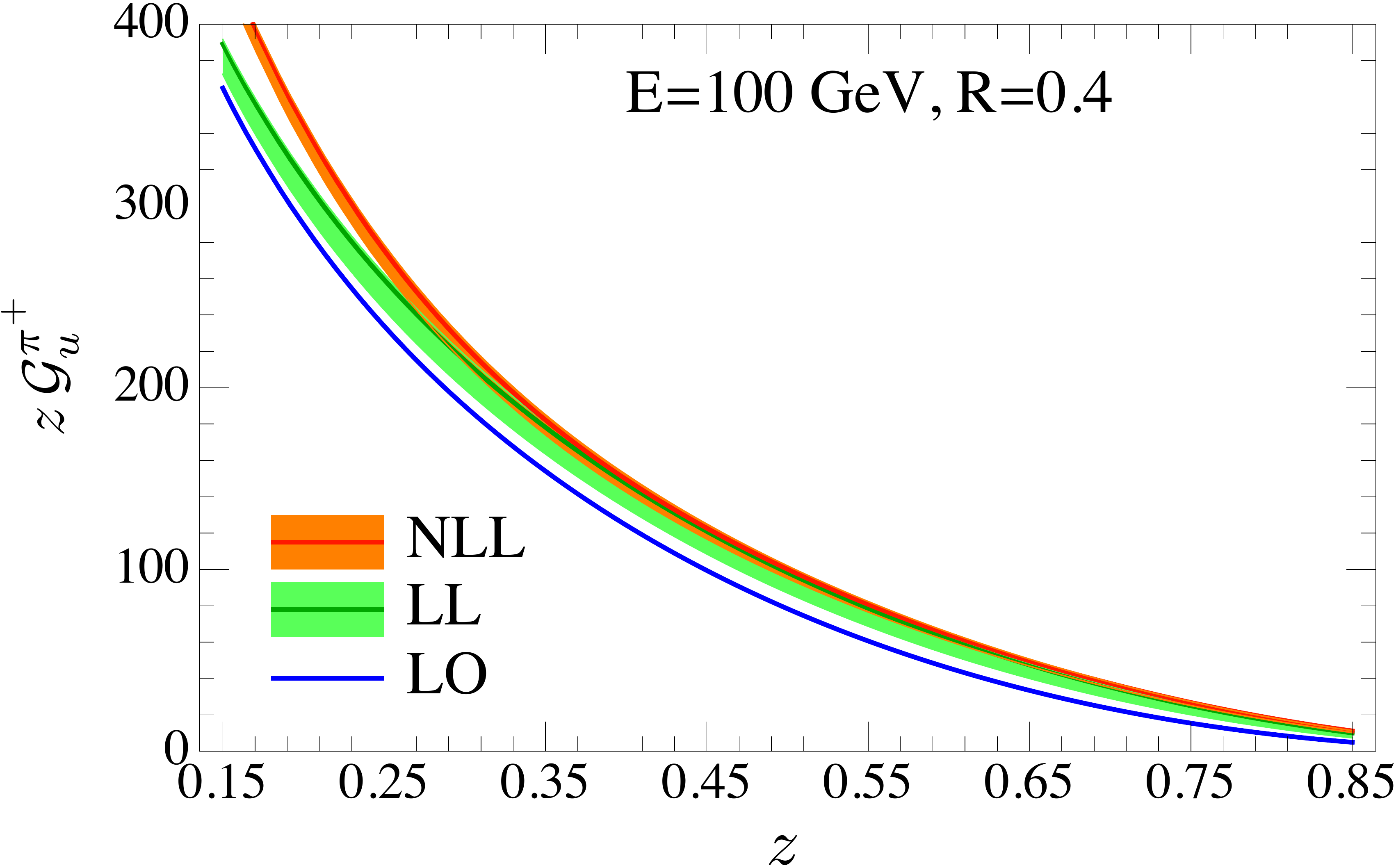}%
\hfill%
\includegraphics[width=0.328\textwidth]{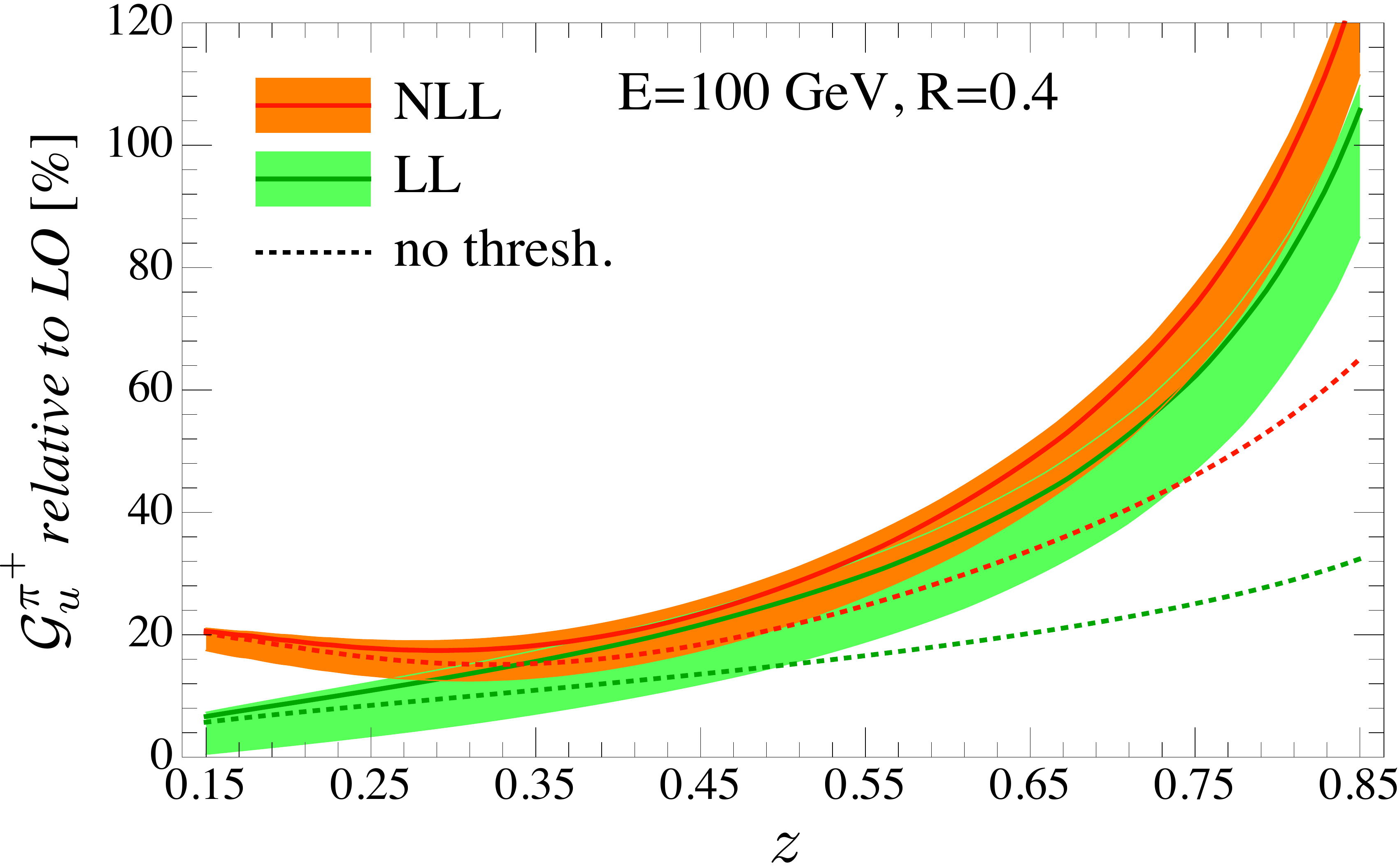}%
\hfill%
\includegraphics[width=0.328\textwidth]{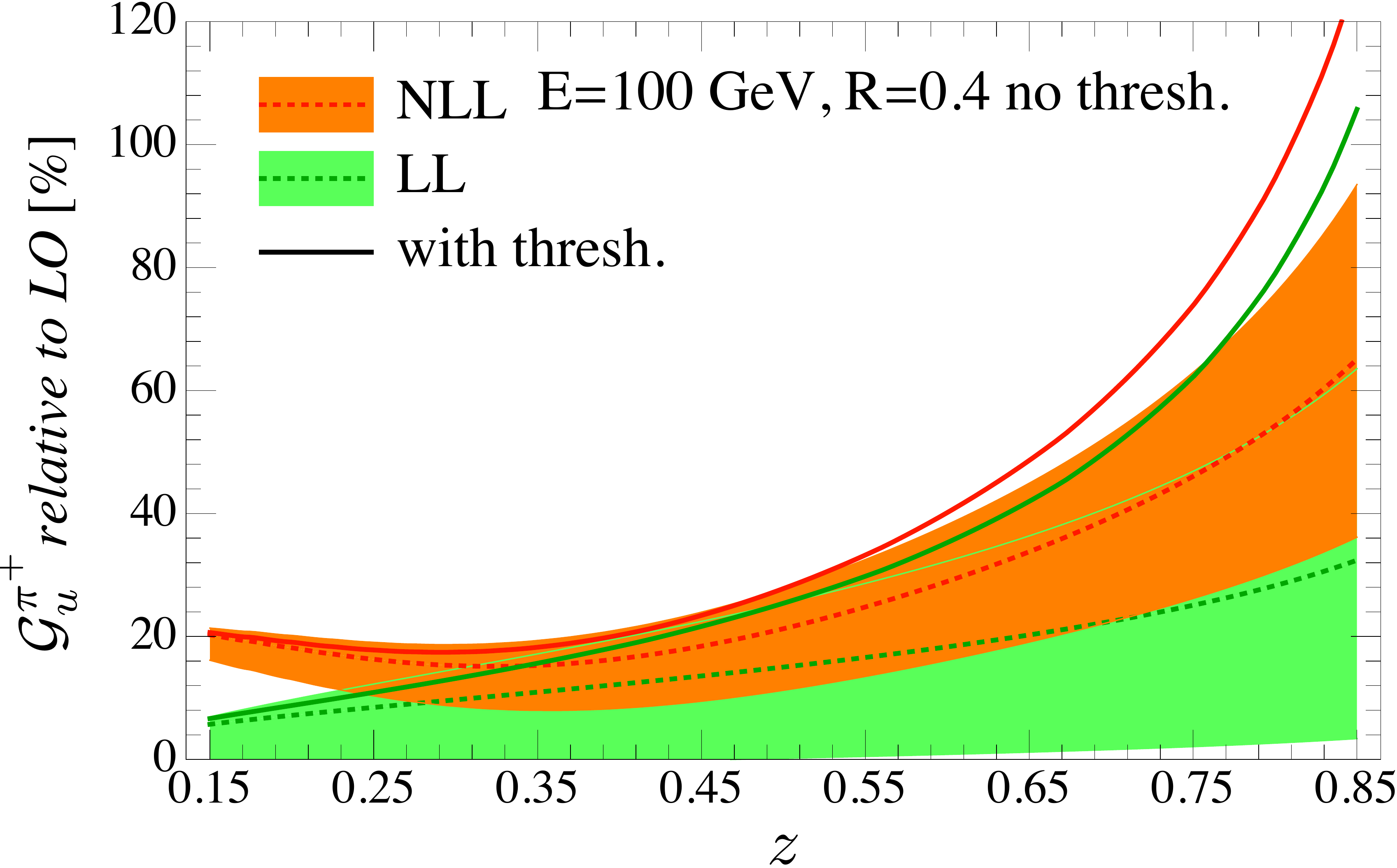}%
\vspace{-2ex}
\caption{\label{fig:u_conv} The FJF for $u \to \pi^+$ at LL and NLL order for $E=100\,$GeV and $R=0.4$. The bands show the perturbative uncertainties from varying the scale in the matching onto $D_j^{\pi^+}$ up and down by factors of two, and then evolving $\cG_u^{\pi^+}$ to $\mu = 2E\,\tan(R/2)$. For reference we include the LO result (where $\mu$ is the jet energy $E$) in the left panel. In the middle panel the same curves and bands relative to this LO result are shown. The dotted lines correspond to the central values without threshold resummation, for which the uncertainties are shown in the right panel. \vspace{-2ex}}
\end{figure*}

\paragraph*{Calculation.}

We now calculate the one-loop matching coefficients for the cone FJFs $\cG_i^h(R,z,\mu)$ onto FFs. At one-loop the cone restriction is equivalent to 
 \begin{equation}
   s \leq \text{min}\Big(\frac{z}{1-z},\frac{1-z}{z}\Big)\, 4E^2 \tan^2 (R/2)
\,,\end{equation}
 where $s$ is the invariant mass of the jet, $z$ the fragmentation variable, $E$ the jet energy and $R$ the cone size. [Note that $E$ and $R$ appear in the combination $E^2 \tan^2 (R/2)$ which is invariant under boosts along the jet axis.] We may therefore obtain the matching coefficients from the bare results for the standard FJFs in Ref.~\cite{Jain:2011xz}. We find, using the $\overline{\text{MS}}$ scheme,
\begin{align} \label{eq:Jij}
   \frac{\cJ_{qq}(E,R,z,\mu)}{2(2\pi)^3} &= \de(1-z) + \frac{\al_s C_F}{\pi} \bigg[\de(1-z) \Big(L^2 - \frac{\pi^2}{24}\Big) 
   \nn \\ & \hspace{19ex}   
   + P_{qq}(z) L + \widehat \cJ_{qq}(z) \bigg]
   \,,\nn \\
   \frac{\cJ_{qg}(E,R,z,\mu)}{2(2\pi)^3} &= \frac{\al_s C_F}{\pi} \big[P_{gq}(z) L + \widehat \cJ_{qg}(z) \big]
   \,,\nn \\
   \frac{\cJ_{gg}(E,R,z,\mu)}{2(2\pi)^3} &= \de(1-z) + \frac{\al_s C_A}{\pi} \bigg[\de(1-z) \Big(L^2 - \frac{\pi^2}{24}\Big) 
   \nn \\ & \hspace{19ex}
   + P_{gg}(z) L + \widehat \cJ_{gg}(z) \bigg]
   \,,\nn \\
   \frac{\cJ_{gq}(E,R,z,\mu)}{2(2\pi)^3} &= \frac{\al_s T_F}{\pi} \big[P_{qg}(z) L + \widehat \cJ_{gq}(z) \big]
\,,\end{align}
with the splitting functions in the convention of Eq.~(3.7) in Ref.~\cite{Jain:2011iu}. Anti-quarks have the same coefficients as quarks, and $\cJ_{q\bar q}$ and $\cJ_{qq'}$ only start at two-loop order. In \eq{Jij}
\begin{equation} \label{eq:scale}
  L = \ln \frac{2 E\tan (R/2)}{\mu}
\end{equation}
and
\begin{align} 
  \widehat \cJ_{qq}(z) &= \frac{1}{2} (1- z) + \begin{cases}
  P_{qq}(z) \ln z & z\leq \frac{1}{2} \\
  (1+z^2) \big(\frac{\ln(1-z)}{1-z}\big)_{\!+}  & z\geq \frac{1}{2} \\
  \end{cases}
  \,, \nn \\
  \widehat \cJ_{qg}(z) &= \frac{z}{2} + P_{gq}(z) \begin{cases}
  \ln z & z\leq \frac{1}{2} \\
  \ln(1-z) & z\geq \frac{1}{2} \\
  \end{cases}
  \,, \nn \\  
    \widehat \cJ_{gg}(z) &= \begin{cases}
  P_{gg}(z) \ln z & z\leq \frac{1}{2} \\
  \frac{2(1-z+z^2)^2}{z} \big(\frac{\ln(1-z)}{1-z}\big)_{\!+}  & z\geq \frac{1}{2} \\
  \end{cases}
  \,, \nn \\  
    \widehat \cJ_{gq}(z) &= z(1-z) + P_{qg}(z) \begin{cases}
  \ln z & z\leq \frac{1}{2} \\
  \ln (1-z)  & z\geq \frac{1}{2} \\
  \end{cases}  
\,.\end{align}

We have cross-checked these results with expressions for the jet functions $J_i$ with a cone restriction in Ref.~\cite{Ellis:2010rwa} (there called ``unmeasured jet functions"). The FJFs $\cG_i$ and jet functions $J_i$ have the same anomalous dimension \cite{Procura:2009vm}, consistent with \eq{fact}, which provides a check on the UV-divergent terms that we encounter in deriving \eq{Jij}. The finite terms were checked with the momentum sum rule \cite{Procura:2009vm,Jain:2011xz}
\begin{equation}
    \sum_j \int_0^1 \df z\, z\, \cJ_{ij}(R,z,\mu) = 2(2\pi)^3\, J_i(R,\mu)
\,,\end{equation}
and a new sum rule
\begin{equation}
   \int_0^1 \df z\, [\cJ_{qq}(R,z,\mu) - \cJ_{q\bar q}(R,z,\mu)] = 2(2\pi)^3\, J_q(R,\mu)
\,.\end{equation}
This new relation follows from quark-number conservation in the perturbative calculation of the FJF $\cG_q^i$.

\begin{figure*}
\includegraphics[width=0.326\textwidth]{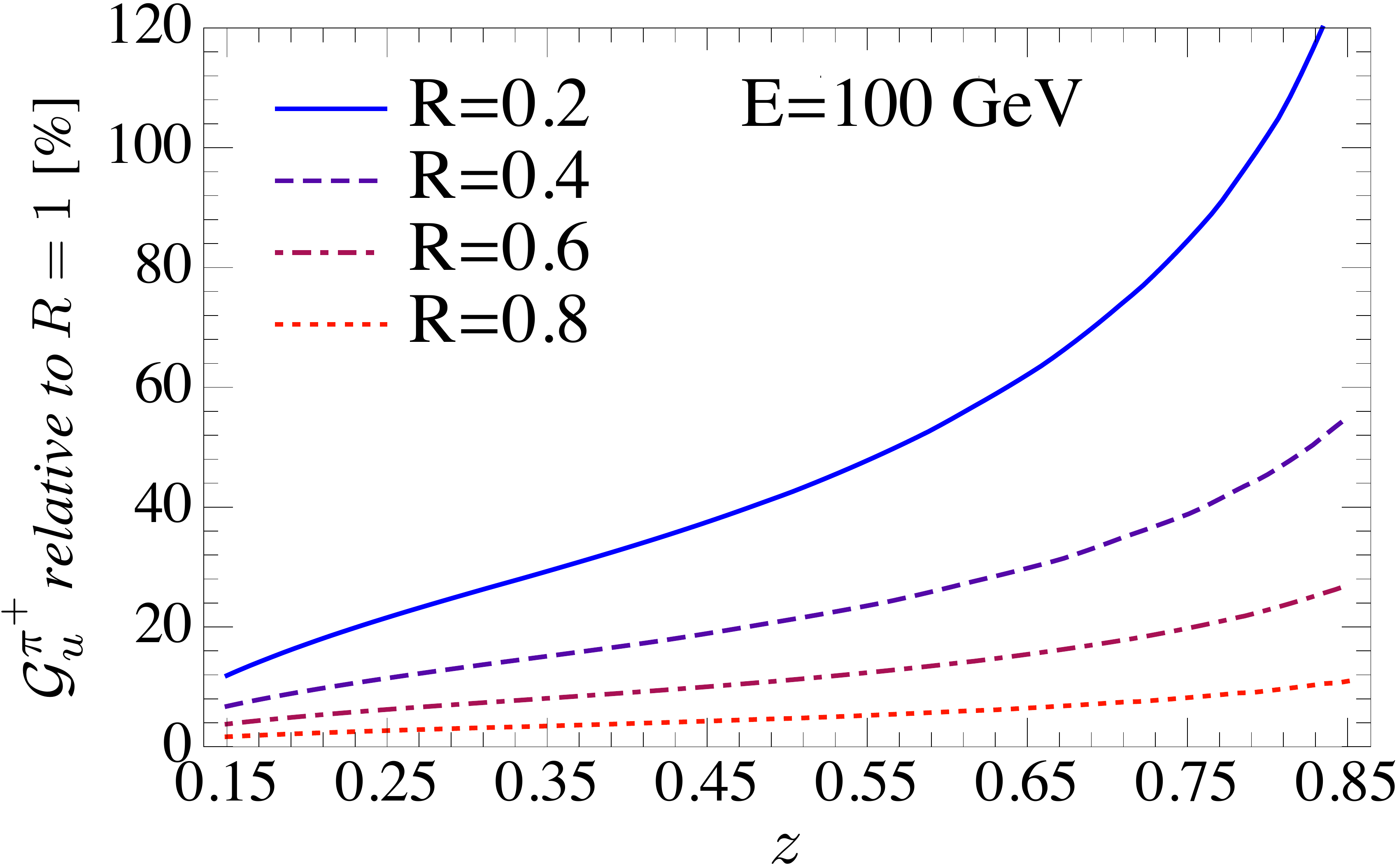}%
\hfill%
\includegraphics[width=0.326\textwidth]{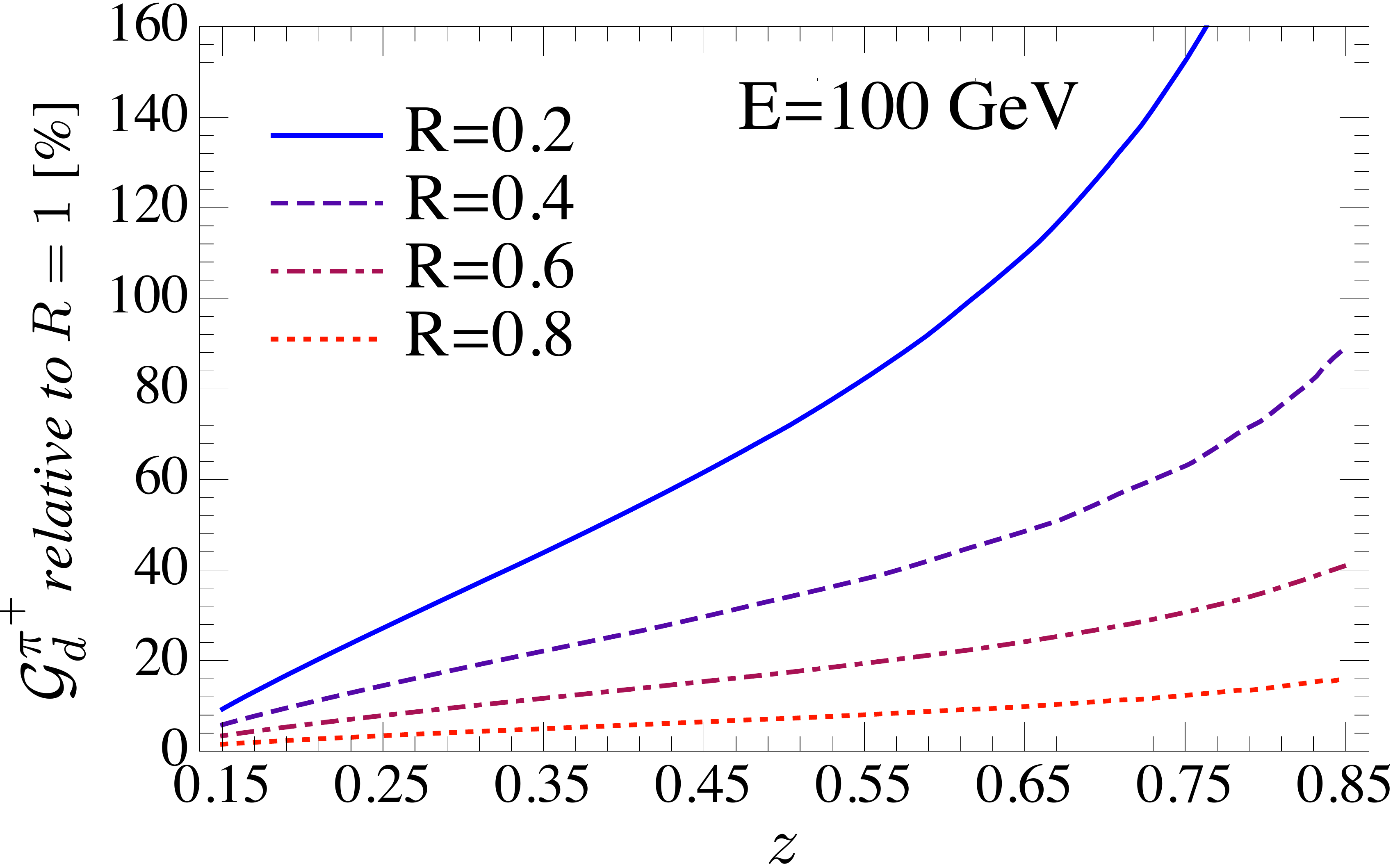}%
\hfill%
\includegraphics[width=0.326\textwidth]{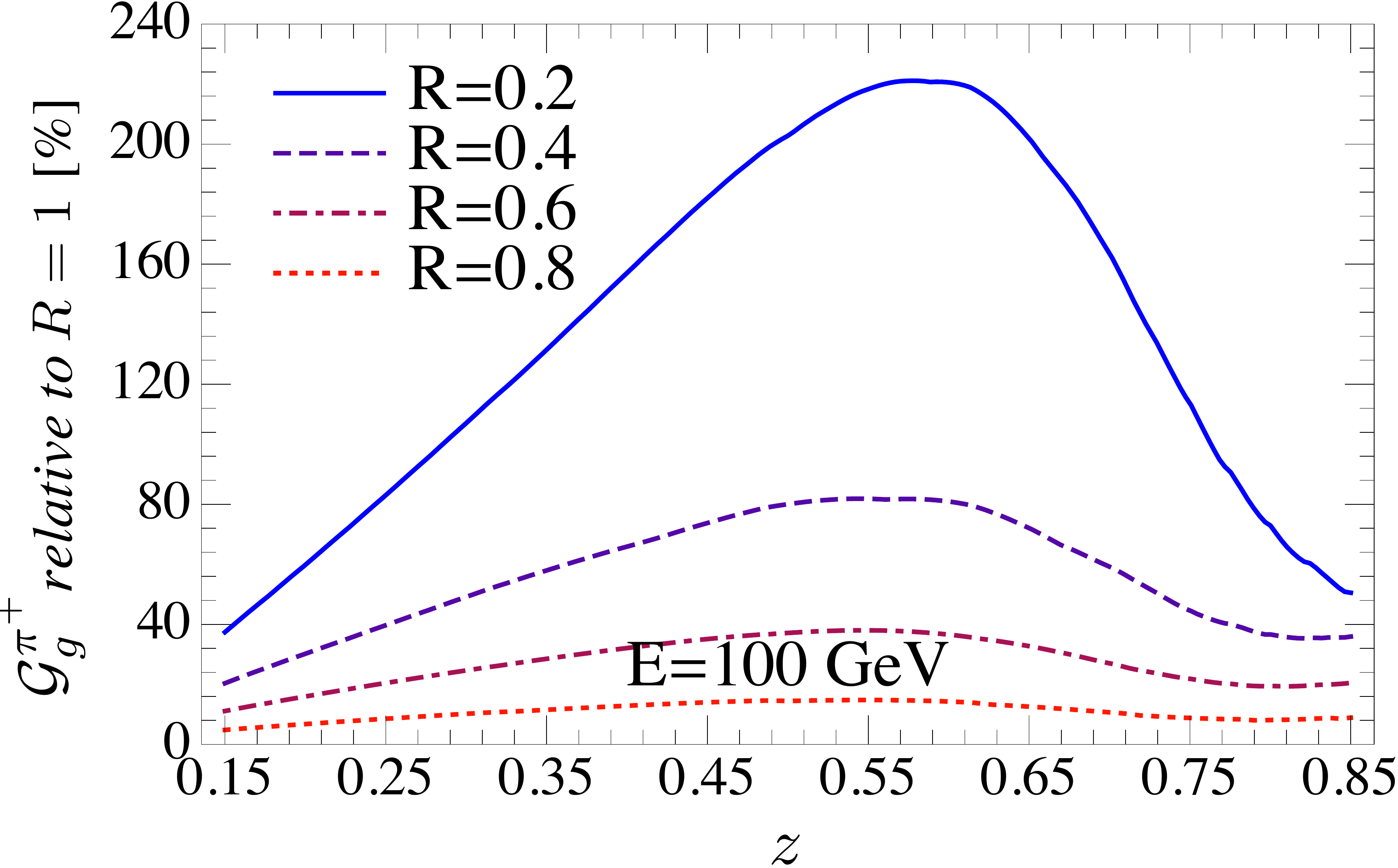}%
\vspace{-2ex}
\caption{\label{fig:rdep} The FJF for $u, d, g \to \pi^+$ is shown at NLL order for $E=100\,$GeV. The curves are for $R=0.2,0.4,0.6,0.8$ and are shown relative to the $R=1$ result.}
\vspace{-2ex}
\end{figure*}

\paragraph*{Threshold Logarithms.}

Closer inspection reveals that $\cJ_{qq}$ and $\cJ_{gg}$ contain large threshold logarithms. For example,
\begin{align} \label{eq:thr1}
 \frac{\cJ_{qq}(E,R,z,\mu)}{2(2\pi)^3} &= \frac{\al_s C_F}{\pi} \bigg[\de(1-z) \Big(L^2 - \frac{\pi^2}{24}\Big) + \frac{2L}{(1-z)}_{\!+} 
 \nn \\ & \hspace{4ex}
 + 2\Big(\frac{\ln(1-z)}{1-z}\Big)_{\!+} + \ord{1-z} \bigg]
\,,\end{align}
which by \eq{GtoD} leads to the large (double) logarithms
\begin{align} \label{eq:thr2}
\frac{\cG_q^h(E,R,z,\mu)}{2(2\pi)^3} &\!=\! \ln^2 \!\Big[\frac{2(1-z) E \tan \frac{R}{2}}{\mu}\Big] D_q^h(z) +\dots
 \end{align}
We may thus sum both the logarithms of $R$ and the threshold logarithms by evaluating the FJF at $\mu = 2(1-z) E \tan (R/2)$. (In the plots we evolve the FJF to $\mu = 2 E \tan (R/2)$, to remove the $z$ dependence from this scale.) \fig{u_conv} shows the improved convergence of perturbation theory arising from threshold resummation, which will be discussed more extensively in the next section. We stress that we do not need to perform a threshold expansion; therefore our results are also valid away from the threshold region.

This improvement holds to all orders in perturbation theory, since we find that the logarithms of $R$ and $1-z$ are tied together through the anomalous dimension of $\cJ_{qq}$ \footnote{This requires the all-orders structure of the jet function anomalous dimension, which is known for hemisphere invariant mass jets, but not established for jet algorithms.}. Using \eq{GtoD} this can be obtained from the anomalous dimension of the (unmeasured) jet function~\cite{Ellis:2009wj,Ellis:2010rwa} and the FF~\cite{Korchemsky:1992xv},
\begin{align}
 \gamma_{\cJ_{qq}} &= -2 \Ga_q \Big[L\, \de(1-z) + \frac{1}{(1-z)}_{\!+}\Big] 
  \nn \\ & \quad 
  + \tilde \ga_\cJ \de(1-z) + \ord{1-z}
\,.\end{align}
Here $\Ga_q = \al_s C_F/\pi + \ord{\al_s^2}$ is the (quark) cusp anomalous dimension. The non-cusp anomalous dimension $\tilde \ga_\cJ$ only starts at two loops, which is reflected in the absence of single logarithms in \eqs{thr1}{thr2}. In moment space one can see explicitly how the logarithm of $R$ and the $1/(1-z)_+$ combine, since the $N$th moment of $1/(1-z)_+$ is $-\ln N -\ga_E$. 

The above discussion applies to $\cJ_{gg}$ as well. By contrast, the off-diagonal $\cJ_{qg}$ and $\cJ_{gq}$ do not contain threshold logarithms and their contribution should thus be evaluated at the appropriate scale $\mu = 2E \tan (R/2)$.

Our new joint resummation can also be used in the standard FJFs $\cG_i^h(s,z,\mu)$ of Refs.~\cite{Procura:2009vm,Liu:2010ng,Jain:2011xz}. Here the scale choice $\mu^2 = (1-z)s$ sums both the logarithms of $s$ and $1-z$. We have checked that this significantly improves the convergence of perturbation theory at large $z$.

Similarly, for initial-state jets described by beam functions $B_i(t,x,\mu)$~\cite{beamf}, we find that the logarithms of $t$ and $1-x$ can be simultaneously resummed by evaluating the flavor-diagonal terms for matching beam functions onto PDFs at the scale $\mu^2 = (1-x)t$. At the LHC, the momentum fraction $x$ is typically small, suggesting that threshold resummation is less important than in the case studied here. However, it has been argued that the logarithms of $1-x$ are dynamically enhanced through the shape of the PDFs~\cite{thr}. A detailed study is left for future work.

\begin{table}[b]
  \centering
  \begin{tabular}{l | c c c c}
  \hline \hline
  & $\cJ_{ij}$ & $\tilde \gamma$ & $\Gamma$ & $\beta$  \\ \hline
  LO & $0$-loop & - & - & $1$-loop \\
  NLO & $1$-loop & - & - & $2$-loop \\
  LL & $0$-loop & - & $1$-loop & $1$-loop\\  
  NLL & $1$-loop & $1$-loop & $2$-loop & $2$-loop\\
  \hline\hline
  \end{tabular}
\caption{Order counting for the matching coefficients, the non-cusp and cusp anomalous dimension, and the $\al_s$ running.\vspace{-1ex}}
\label{tab:counting}
\end{table}

\newpage

\paragraph*{Results.}

We now show some numerical results for fragmentation inside a cone jet. Our order counting is shown in Table \ref{tab:counting}. As input we use the HKNS FFs~\cite{Hirai:2007cx} at LO or NLO order and their value for $\al_s$. 

We first consider a fixed cone $R=0.4$ and study the convergence of our resummed result. \fig{u_conv} shows the central values and perturbative uncertainties for the $u \to \pi^+$ FJF at LL and NLL order. In the middle and right panels these results relative to the LO curve are shown, making the separation between the central curves and the size of the uncertainties more visible. As expected, the cone restriction suppresses the small $z$ region because at large $z$ the hadron carries off most of the energy and there is not much additional radiation that would be subject to the cone restriction. As is clear from comparing the dotted with the solid curves, the effect of threshold resummation becomes important at $z \gtrsim 0.5$. Including threshold resummation improves the convergence and leads to smaller uncertainties. At small $z$ the difference between LL and NLL is not fully captured by the uncertainty band. However, in this region the uncertainties on the FFs (which are not shown here) dominate.

We have studied similar plots for $d\to \pi^+$ and $g\to \pi^+$. Here the convergence is not as good when the dominant mixing contribution involving $D_u^{\pi^+}$ (and $D_{\bar d}^{\pi^+}$) turns on. For $g \to \pi^+$ and $d\to \pi^+$ this happens at NLO and NNLO, respectively.

In \fig{rdep} we study the effect of varying the cone size $R$. We stress that the normalization of the curves for different $R$ should not be compared because other ingredients in the factorization theorem affect the normalization of the cross section (the shape in $z$ is entirely determined by $\cG$). For $u\to\pi^+$ and $d\to\pi^+$ the small $z$ region gets suppressed relative to the large $z$ region as one reduces the cone size, as expected. For $g\to\pi^+$, the bump-shaped enhancement at intermediate $z$ is mainly due to the $\cJ_{gq}$ contribution. 

\paragraph*{Conclusions.}

Our framework leads to a reliable description of fragmentation within an identified jet, 
providing the tools for a more exclusive study of fragmentation. This can be used to reduce the background from other processes, to help identify underlying partonic structures and enables novel tests of the universality of FFs. 
Our setup provides accurate analytical predictions, where large logarithms are properly summed, for single hadron spectra in a jet which can be used to tune Monte Carlo event generators. This will become more and more relevant as the determination of the fragmentation functions becomes more accurate.
We have for the first time established the connection between the resummation of logarithms for exclusive jet production and threshold resummation, and have shown the numerical importance of their interplay in the case of jet fragmentation.


We thank T.~Becher, S.~Ellis, A.~Jain, C.~Lee, A.~Manohar, J.~Walsh and F.~W\"urthwein for discussions and comments.
M.P.~acknowledges support  by the ``Innovations- und Kooperationsprojekt C-13'' of the Schweizerische Universit\"atskonferenz SUK/CRUS and by the Swiss National Science Foundation. 
W.W.~is supported by DOE grant DE-FG02-90ER40546. 


\end{document}